\documentclass{article}

\usepackage[preprint,nonatbib]{neurips_2022}

\usepackage[utf8]{inputenc}
\usepackage[T1]{fontenc}   
\usepackage{hyperref}      
\usepackage{url}           
\usepackage{booktabs}      
\usepackage{amsfonts}      
\usepackage{amsmath}
\usepackage{nicefrac}      
\usepackage{microtype}     
\usepackage{xcolor}        
\usepackage{graphicx}
\usepackage{wrapfig}
\usepackage{algorithm}
\usepackage{algorithmicx}
\usepackage{algpseudocode}
\usepackage{amssymb}

\title{HCMD-{\it zero}: Learning Value Aligned Mechanisms from Data}

\author{Jan Balaguer, Raphael K\"oster, Ari Weinstein, Lucy Campbell-Gillingham,\\
\textbf{Christopher Summerfield, Matt Botvinick \& Andrea Tacchetti}\\
DeepMind, UK\\
\texttt{\{jua, rkoster, ariweinstein, lcgillingham}\\
\texttt{csummerfield, botvinick, atacchet\}@deepmind.com}\\}

\begin{document}

\maketitle

\begin{abstract}
Artificial learning agents are mediating a larger and larger number of interactions among humans, firms, and organizations, and the intersection between mechanism design and machine learning has been heavily investigated in recent years. However, mechanism design methods often make strong assumptions on how participants behave (e.g. rationality), on the kind of knowledge designers have access to a priori (e.g. access to strong baseline mechanisms), or on what the goal of the mechanism should be (e.g. total welfare).
Here we introduce HCMD-{\it zero}, a general purpose method to construct mechanisms making none of these three assumptions. HCMD-{\it zero} learns to mediate interactions among participants and adjusts the mechanism parameters to make itself more likely to be preferred by participants. It does so by remaining engaged in an electoral contest with copies of itself, thereby accessing direct feedback from participants. We test our method on a stylized resource allocation game that highlights the tension between productivity, equality and the temptation to free ride. HCMD-{\it zero} produces a mechanism that is preferred by human participants over a strong baseline, it does so automatically, without requiring prior knowledge, and using human behavioral trajectories sparingly and effectively. 
Our analysis shows HCMD-{\it zero} consistently makes the mechanism policy more and more likely to be preferred by human participants over the course of training, and that it results in a mechanism with an interpretable and intuitive policy.
\end{abstract}

\section{Introduction}
\begin{wrapfigure}{R}{6cm}
    \centering
    \includegraphics[width=6cm]{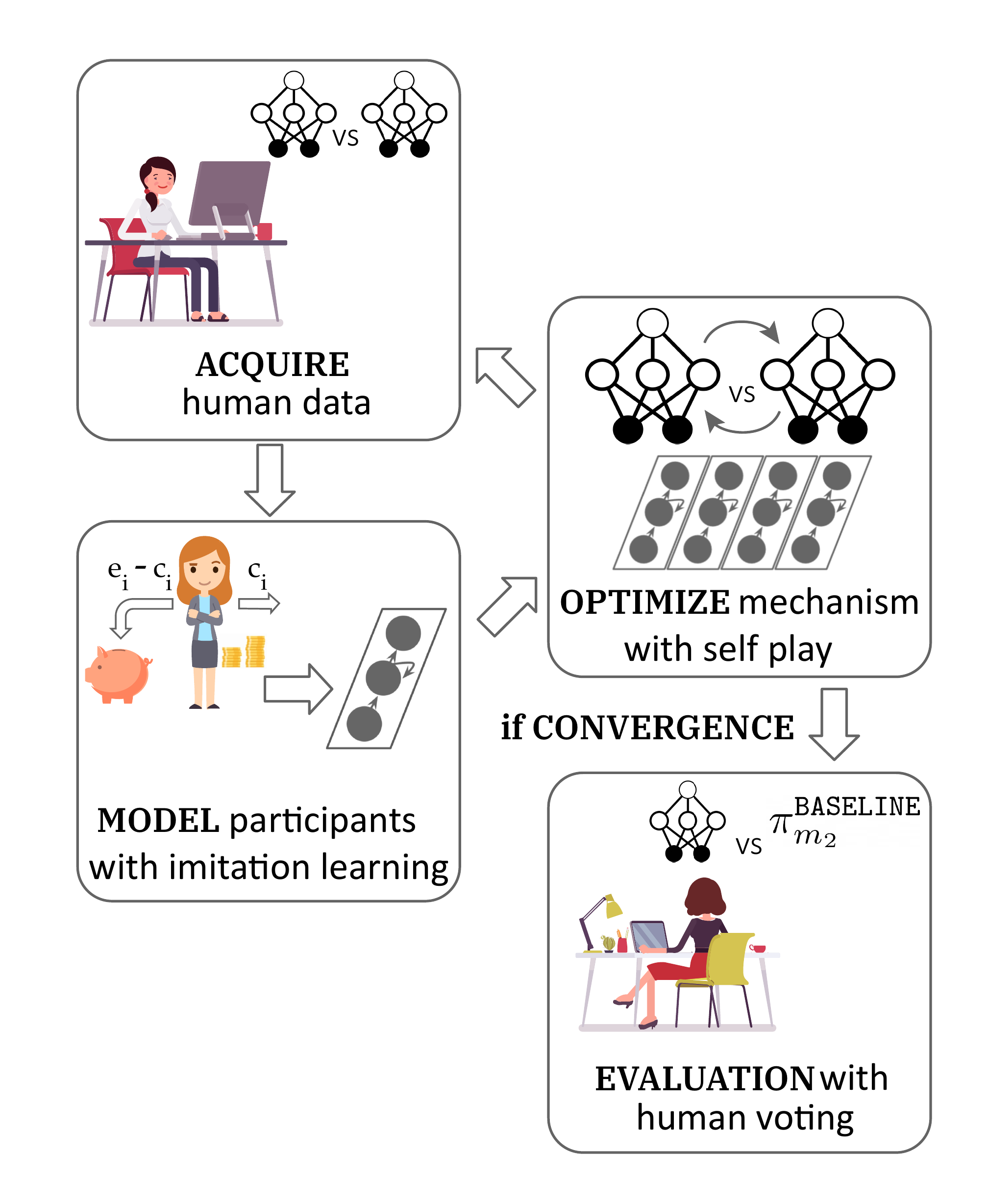}
    \caption{\small Illustration of HCMD-\textit{zero} training and evaluation. HCMD-\textit{zero} starts by acquiring human game play trajectories. The data is then used to model human participants behavior. The mechanism network is then trained in self-play using the most recent human participants models in the environment loop. After HCMD-\textit{zero} has converged the mechanism is evaluated with new human participants and against a baseline mechanism.}\label{fig:hcmd-cartoon}
\end{wrapfigure} 

Artificial learning agents are beginning to play a central role in our institutions. From social networks, to investment management, and traffic routing, an ever growing number of interactions among humans, firms and organizations are mediated by adaptive systems.

While the intersection between mechanism design and machine learning has been heavily investigated in recent years, most methods make strong assumptions on either the behavior and goals of participants (e.g. rationality), or on the kinds of knowledge, baseline mechanisms, or data we have access to before constructing a new mechanism for a given economic interaction. Moreover, often the goal of the mechanism is chosen arbitrarily by the designer, without regard to what participants might prefer (e.g. pure welfare, or a specific equality adjusted metric of welfare).

Here we address these restrictive assumptions and present a general method to design a mechanism that is able to mediate complex economics interactions among human participants, and that is preferred by humans over baseline alternatives. Our method requires no access to alternative mechanisms during training, and it makes no assumptions on the nature of participants' preferences, values or strategies.

The main challenges in our work stem from making no assumptions on what people want, which reveals itself in two ways. The first is what the goal of the mechanism should be. For example, given a complex policy (i.e. a distribution policy in a multi-player economic game) it is not feasible to collect information about what participant's favoured response of the mechanism would be for each possible game state. Inspired by real life democracy, we turn to estimation of preferences over sampled experiences: participants get to sample two mechanisms' policies in two games, and return a vote which one they preferred and want to play with again. This is the reward signal the mechanism will be trained to optimize. \textbf{Voting is used as a quantification of how value-aligned the mechanism is with the population that it is interacting with}. Second, to optimize the mechanism policy in the economic interaction game itself, the mechanism needs a to access a model of human decision making during training. Again, we do not want to make any assumptions about humans' behavior. In particular, we do not want to assume participants payoffs which could be used to estimate their strategy~\cite{lanctot2017unified,balaguer2022good,kim2022influencing}). Similarly, we do not want to assume that voting behavior and contribution behavior in the investment game share the same goals, as preferences over group outcomes and individual outcomes could differ. Therefore, we turn to \textbf{imitation learning, copying human behavior without any assumptions about the goals they may pursue}.

We test our method on a stylized investment game where human participants could earn real money and which is known to stress the tensions between welfare, equality and the temptation to free ride. Our results show that the method presented here was able to construct a complex mechanism policy based on a simple expression of preference; and that this policy is favored by novel participants over a baseline that was previously established as strong in this task. Our analysis shows how the election-against-self curriculum pushes our mechanism towards interpretable mediation schemes with increasingly more pronounced punish / reward regions.

The impact of AI on our institutions is growing rapidly; and as such the intersection of mechanism design and machine learning is receiving considerable attention. Here we show that merging the most basic democratic principle of ``one person, one vote'' with modern machine learning and game theory insights leads to a general method for designing mechanisms that are aligned with the preferences of their constituents.

\section{Related work}

Value alignment and AI safety have been intensely investigated in recent years both from a normative perspective~\cite{gabriel2020artificial}, and from a technical one~\cite{dafoe2020open}; and there is growing support for building participatory systems for AI ethics and governance~\cite{iyad2017society,lee2019webuildai}.

Mechanism design is a sub-field of economics that studies how to design the rules and incentives of multi-agent interactions, so that self-interested participants will prefer certain strategies, often trading off their own welfare for that of the group. The field has a long history to which it is near impossible to do justice, see~\cite{maskin2008mechanism} for a review. More recently mechanism design has been studied from an algorithmic point of view~\cite{conitzer2002complexity,nisan2001algorithmic}, as well as a machine learning one~\cite{dutting2017optimal,manisha2018learning,tacchetti2019neural,koster2022democratic,balaguer2022good,kramar2017should}. Finally, researchers have recently turned their attention to the role that mechanism design can play in our pursuit of social good~\cite{abebe2018mechanism}.

Agent based models (ABMs), where a computer simulation predicts how autonomous agents will adapt to certain environment interventions, has been a tool used by policy makers to design new mechanisms since its inception. ABMs have received renewed attention after the 2008 Economic Crisis~\cite{farmer2009economy,hamill2015agent}.

The problem of building artificial agents that coordinate with human participants starting from ``zero knowledge'' had been investigated both in the computer game setting~\cite{strouse2021collaborating}, and in a simulated economy environment~\cite{zheng2020ai}. Similarly, self-play and assumption free methods have been successfully applied to challenging constant-sum two-player games in recent past~\cite{silver2016mastering,silver2017mastering,vinyals2019grandmaster}.

\section{Methods}

We start this section by introducing the Public Investment Game (PIG) we consider in our experiments so as to provide a grounding example for the exposition of our method. We then introduce the HCMD-\textit{zero} algorithm in very general terms, and finally describe how HCMD-\textit{zero} can be applied to the PIG in particular.

\subsection{Public Investment Game for participants and mechanisms}

\begin{figure}[ht]
    \centering
    \includegraphics[width=0.7\linewidth]{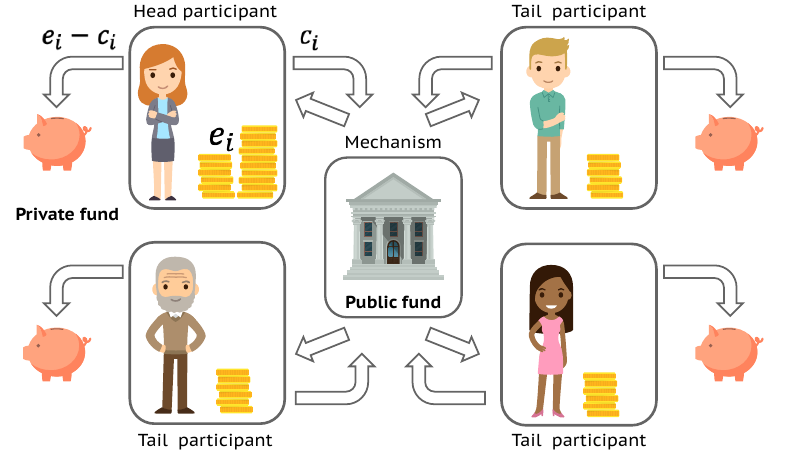}
    \caption{\small Single stage of the Public Investment Game (PIG): 4 participants play over 10 identical rounds. Each round participants receive an endowment (which remains the same for each participant across all rounds). On each round, participants can decide how much to keep private (directly contributing to their monetary payout after the game) or contribute to the public fund. The public fund sums all contributions and multiplies them by 1.6. The public fund is redistributed to the participants on each round according to the policy of the mechanism. The returns from the fund then are added to each participant's monetary payout. Thus, the policy of the mechanism can influence whether the game is a social dilemma and how much initial inequality is redressed by the payouts of the fund. Figure reproduced with permission from \cite{koster2022democratic}.}
    \label{fig:game}
\end{figure}

From the point of view of the 4 participant players, the Public Investment Game unfolds over 3 stages. In Stages 1 and 2 the 4 the participant players interact with two mechanism players $\pi_{m_1}$ and $\pi_{m_2}$ respectively, while in Stage 3, the mechanism player is selected by majority vote as explained below. In particular, the first stage proceeds over 10 identical rounds where, at the beginning of each round, each participant receives an endowment of $e_{i, t}$ ``coins'', with $i=1,\ldots,4$, and  $t=1,\ldots,10$, and decides what fraction of coins $\rho_{i,t}$ they would like to invest in a public fund that grows with a fixed multiplier of $1.6$. A mechanism player $\pi_{m_1}$ then observes $e_{i, t}$ and $\rho_{i,t}$, and determines 4 redistribution weights $w_{i,t} \geq 0$, with $\sum_i w_{i,t} = 1$, according to which the fund is returned in its entirety: each participant receives $w_{i,t}(1.6\times\sum_i \rho_{i,t}e_{i,t})$, and the game moves on to the next round. The second stage is identical to the first, with the exception that participants interact with mechanism player $\pi_{m_2}$ instead of $\pi_{m_1}$. After the second stage, each participant casts a vote on which mechanism they would like to re-experience in the third stage, which is again identical to the first 2, except that the mechanism player is decided by majority vote.

In our experiments, at the end of each stage, each human participant collected a monetary reward proportional to the funds they received from the public investment fund, and the endowments they decided not to contribute $R_i = \sum_t r_{i,t} = \sum_t w_{i,t}(1.6\times\sum_i \rho_{i,t}e_{i,t}) + (1-\rho_{i,t})e_{i,t}$. Participants thus had ``skin in the game'' when reporting which mechanism they preferred, as their vote could decide what mechanism they encountered in the third and final stage.

From the vantage point of the participants, and for fixed mechanism policies, the (curried) PIG is a general-sum 4 player game with the redistribution decisions of each mechanism folded in the game transition kernel. Participants observe all game events but they are not informed about the nature of the mechanism policy (i.e. they know the mechanism changes in each stage, but they don't know what mechanisms do in response to various participants' joint behaviors).

From the point of view of the mechanism player, and for fixed participants policies, the (curried) PIG is a 2 player constant-sum symmetric game. First, the two candidate mechanisms face independent sequential decision making tasks (one in each of the two initial stages), with states coinciding with endowments, contributions, and redistribution histories, and actions coinciding with redistribution weights. Second, the two mechanisms collect a payoff based on the number of votes cast in their favor. Similarly to what happens with participants, the transition kernel from their vantage point implements the PIG game dynamics, as well as the participants contributing and voting behaviors.

\subsubsection{Qualitative analysis of the game}

In this paragraph we surface the broad strategic traits of a single stage of the PIG by walking through the effects \textit{on the behavior of rational participants} of two extreme choices for a mechanism policy. 1) For a trivial mechanism redistributing equally to every player at every step, and irrespective of their contribution ('Strict Egalitarian'), the curried PIG for the participants is a clear social dilemma: each participant wishes for everyone to contribute generously to the public fund while having no incentives to contribute themselves. The only Nash equilibrium is to defect and contribute nothing at all. 2) A mechanism that returns to each participant proportionally to their contributions, on the other hand, sets up each stage of the curried PIG so that rational participants are incentivized to contribute everything.

However, the mechanism we wish to construct aims to maximize getting voted for by participants, and this adds significant complications. Participants start each round with unequal endowments (one participant having 10 coins to give, the others 2, 4, 6, 8 or 10). Mechanisms that redistribute wealth to the 3 participants with lower endowments, stand to gain more votes from participants maximizing their own welfare. Importantly, if a mechanism redistributes too much wealth, it will discourage the participant with 10 coins to contribute, reducing the amount of wealth that can be redistributed. Even assuming that all participants want to maximize their own welfare, the mechanism policy needs to take into account how fast and well participants learn and adjust their contributions to the given incentives. Additionally, it is not clear whether participants actually aim to maximize their welfare, or if their vote and contributing behavior is also informed by other principles (e.g. fairness).

\subsection{HCMD-\textit{zero}}

In general terms, we consider the problem of constructing a mechanism policy that mediates economic interactions among human participants, and that is preferred by these same human participants over a strong baseline alternative that is not known at training time. Additionally, our method assumes that interactions with humans are hard or expensive to obtain, and thus online learning from these interactions is infeasible.

Formally, we seek to maximize the payoff of a mechanism player $\pi_{m_1}$ interacting with a further mechanism player $\pi_{m_2}^{\text{\texttt{BASELINE}}}$ (the baseline alternative mechanism), and $N$ human participant players $(h_p^1,\ldots,h_p^N) \triangleq \bar{h}_p$ in an extensive form game $\mathcal{G}$($\pi_{m_1}, \pi_{m_2}^{\text{\texttt{BASELINE}}}, \bar{h}_p)$. In particular, we assume that the participants' side of the payoff is unknown, that for any joint strategy adopted by the participant players, the interaction between the two mechanisms is symmetric and constant sum, and that the policy of the baseline mechanism $\pi_{m_2}^{\text{\texttt{BASELINE}}}$ is not known at training time. It can be easily verified that the PIG satisfies these assumptions.

\paragraph{Training:} The focal point of our method is that since we do not get to know what human participants are trying to maximize, nor have any idea how they might go about maximizing it, we cannot use standard techniques to anticipate their strategies by constructing or approximating best responses. Instead we alternate between using imitation learning to predict participants' behavior, with improving mechanism players' strategies using self-play.

We start by defining a suitable parameterization for the policy of both mechanism players ($\mathcal{G}$ is symmetric for the two mechanisms) and models of the participants behaviors $f_m$ and $f_p$ respectively, mapping the respective observations to the respective action spaces (e.g. endowment and contribution histories to redistributions, or endowments and contribution histories to contributions), and let $\theta_m \in \mathbb{R}^{d_m}$, and $\bar{\theta}_p = (\theta_p^1 \in \mathbb{R}^{d_p^1},\ldots,\theta_p^N \in \mathbb{R}^{d_p^N})$ be parameter vectors for $f_m$ and $f_p$ respectively.  We then construct games $\mathcal{G}_{\theta_m^*, \theta_m^*} = \mathcal{G}(f_m(\theta_m^*), f_m(\theta_m^*), \cdot)$ and $\mathcal{G}^{\bar{\theta}_p^*} = \mathcal{G}(\cdot,\cdot,f_p(\bar{\theta}_p^*))$ by fixing the parameters of both mechanisms to $\theta_m^*$, and participants to $\bar{\theta}_p^*$, and folding the resulting strategies in the corresponding transition tables. In other words, we consider the game from the vantage point of the participants players exclusively in $\mathcal{G}_{\theta_m^*, \theta_m^*}$, and mechanism players exclusively in $\mathcal{G}^{\bar{\theta}_p^*}$; this step is usually referred to as \textit{currying} \cite{balduzzi2019openended}.

HCMD-\textit{zero} (see Algorithm~\ref{algo:amor}) then proceeds, by alternating behavioral data acquisition and modeling of human participants game-play on $\mathcal{G}_{\theta_m^s}$ (\texttt{ACQUIRE} and \texttt{MODEL} steps, with $\bar{h}_p^s$ denoting the human participants recruited at iteration $s$), with the improvement of the mechanism player's policy using a standard self-play loop on $\mathcal{G}^{\bar{\theta}_p^s}$ (\texttt{OPTIMIZE} step). The Algorithm terminates when the \texttt{OPTIMIZE} step fails to yield an improvement over previous iterations (\texttt{CONVERGENCE} step).

Note that since the policy for the mechanism player of interest is trained in self-play on $\mathcal{G}^{\bar{\theta}_p^*}$, the second mechanism player $\pi_{m_2}^{\text{\texttt{BASELINE}}}$ (i.e. the strong baseline we will be tested against) does not appear in the algorithm, and thus can remain unknown at training time. HCMD-\textit{zero} is outlined in Algorithm~\ref{algo:amor} and illustrated in Fig.~\ref{fig:hcmd-cartoon}, with each step further expanded in the following sections.

\begin{algorithm}[ht]
    \begin{algorithmic}
        \Require $\mathcal{G}, \theta_m^0$
        \State $s \gets 1$
        \While{True}
            \State $D_s \gets \text{\texttt{ACQUIRE}}(\mathcal{G}_{\theta_m^{s-1}, \theta_m^{s-1}},\bar{h}_{p}^s)$
            \State $\bar{\theta}_p^s \gets \text{\texttt{MODEL}}(D_1,\ldots,D_s)$
            \State $\theta_m^{s} \gets \text{\texttt{OPTIMIZE}}(\mathcal{G}^{\bar{\theta}_p^s},\theta_m^{s-1})$
            \If{$\text{\texttt{CONVERGENCE}}(\mathcal{G}^{\bar{\theta}_p^s}, \theta_m^0,\ldots,\theta_m^s)$}
                \State return $\theta_m^s$
            \EndIf
            \State $s \gets s + 1$
        \EndWhile
    \end{algorithmic}
    \caption{\small Overview of the HCMD-\textit{zero} algorithm.}\label{algo:amor}
\end{algorithm}

 \paragraph{Evaluation:} We evaluate the performance of the policy parameters by recruiting a new cohort of human participants $\bar{h}^{\text{\texttt{EVAL}}}_p$ and letting them interact with both the mechanism we trained and the baseline mechanism in $\mathcal{G}(f_m(\theta_m^s), \pi_{m_2}^{\text{\texttt{BASELINE}}}, \bar{h}^{\text{\texttt{EVAL}}}_p)$ and reporting the payoff obtained by $f_m(\theta_m^s)$.

\subsection{HCMD-\textit{zero} applied to the PIG}
\subsubsection{\texttt{ACQUIRE} step: Data Acquisition }

The goal of the \texttt{ACQUIRE} step is to gather the data-set $D_s$ of human participants gameplay on $\mathcal{G}_{\theta_m^s,\theta_m^s}$.

We used a crowd-sourcing platform to acquire contributing and voting behavior data from human participants ($n=1656$). All participants gave informed consent to participate in the experiment and the study was approved by an internal review board. Participants pay had a minimum pay and a bonus depending on game performance, averaging \$18 an hour (the total costs of the study were approximately \$40,000). During each iteration $s$, groups of 4 human participants completed the two initial stages of the PIG game interacting with a mechanism player endowed with the most recent parameters (see Sec.~\ref{sec:optimize} for details on the mechanism player), and voted for the episode they preferred (constituting $D_s$). Since the mechanism and the conditions (e.g. the endowment) are identical in both stages, the only difference between them is driven by the randomness in human behavior.

\begin{table*}[h]
    \centering
    \begin{tabular}{l | rrrrrrr}
        \toprule
        \midrule
        Iteration & $s=1$ & $s=2$ & $s=3$ & $s=4$ & $s=5$ & $s=6$ & $s=7$\\ 
        \midrule        
        Groups & 73 & 45 & 51& 101 & 53 & 49 & 42\\
        Contrib. Linear size & 8 & 8 & 8 & 32 & 32 & 32 & 32\\
        Contrib. LSTM size & 4 & 4 & 4 & 8 & 8 & 8 & 8\\
         \midrule
         \bottomrule
    \end{tabular}
    \caption{\small Amount of data collected and modeling hyper-parameters for each iteration.}\label{tab:data-iters}
\end{table*}

\subsubsection{\texttt{MODEL} step: Model participants}

The goal of the \texttt{MODEL} step for iteration $s$, is to create a model $f_p(\bar{\theta}_p^s)$ of the participant players to predict human contributions and votes from all data-sets $D_{1}, \ldots, D_{s}$ collected thus far.

In our experiments, the contribution model was a neural network similar to that in~\cite{koster2022democratic}, which takes as input each participant's normalized endowments and contributions: $e_{i,t} / 10$ and $c_{i,t} / 10$, as well as each participant's fractional contribution $\rho_{i,t}=c_{i,t} / e_{i,t}$, and outputs the log-likelihood of contributing $0, 1, \ldots,10$ coins ($10$ coins being the maximum endowment). The network is applied independently for each participant and composed of an input linear layer, a LSTM, and an output linear layer. The contribution model was trained to minimize group-wise cross-entropy between predicted and actual contributions.

The votes model is a simple linear layer, which we apply independently for each participant, that takes in the flattened observations from a single episode (10 rounds $\times$ 3 endowment/contribution/payout $\times$ 4 participants) and produces a single output, which can be interpreted as the log-likelihood of voting for the current episode. The same linear layer is applied to both episodes, and a softmax normalization produces the final probabilities. We train this network to minimize group-wise cross-entropy between predicted and actual votes, with an additional $l^2$ regularization loss of the linear layer parameters.

Since the amount of data available increases with every iteration, hyper-parameters must be adjusted each time. In our experiments, we tuned the $l^2$ regularization and network size using cross validation with a random 70\%-30\% train/eval split. We reconstituted the original data-set for training (see Tab.~\ref{tab:data-iters} for details).

\subsubsection{\texttt{OPTIMIZE} step: Train mechanism}\label{sec:optimize}

Similarly to~\cite{koster2022democratic} we use a feed-forward Graph Network~\cite{battaglia2018relational} to construct the parameterization of the mechanism policy $f_m$. In particular, in our experiments, $f_m$ took as as input the current endowment and contribution from each participant (as nodes of a fully connected graph) and output deterministic redistribution weights.
 
During the \texttt{OPTIMIZE} step, we update the mechanism policy parameters $\theta_m^s \rightarrow \theta_m^{s+1}$ with self-play on the curried game $\mathcal{G}^{\bar{\theta}_p^s}$ constructed using the most current participants models $f_p(\bar{\theta}_p^s)$.

Specifically, we trained the mechanism by approximating the mechanism player's policy gradient through a bespoke low variance estimator based on Stochastic Computation Graphs~\cite{schulman2016gradient} that exploits the differentiable structure of the PIG while accounting for the stochastic nature of the participant model's contributions (similar to~\cite{koster2022democratic}), enabling us to run our training within a few hours on a single GPU P100 machines. We note that while this choice is suitable for our setup, the learning rule can be replaced by any Reinforcement Learning technique that fits the problem at hand. During training, we used batches of $1000$ games equally split among the endowment condition we considered: $[10, 2, 2, 2], [10, 4, 4, 4], [10, 6, 6, 6], [10, 8, 8, 8]$ and $[10, 10, 10, 10]$ coind. The mechanism's policy was trained using an ADAM optimizer with learning rate $4\mathrm{e}{-5}$. Finally, we fixed the number of gradient updates to $2000$ for intermediate iterations and $10000$ for the final one. This choice warrants a brief discussion: there is a trade off between how aggressively we require our participants model $p_s$ to extrapolate beyond its training distribution (recall that training data was collected using mechanism parameters $\theta_m^0 \ldots \theta_m^{s-1}$) and how many total iterations (and thus data collection steps) we prescribe.

\subsection{\texttt{CONVERGENCE} step: Determine when to stop}

\begin{figure*}[h]
    \centering
    \includegraphics[width=.8\linewidth]{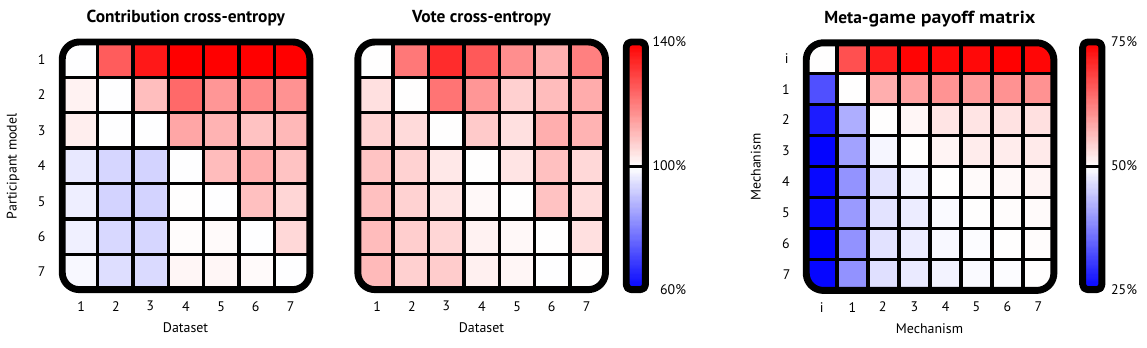}
    \caption{\small Monitoring performance of participant model and mechanism across training iterations. Left panel: contribution and vote cross-entropy (ratio normalized by the diagonal) of participant models $\bar{\theta}_p^1 \ldots \bar{\theta}_p^7$ across data-sets $D_1 \ldots D_7$. Participant models made better predictions on data-sets acquired in earlier iterations. Right panel: payoff matrix from the meta-game, where pairs of mechanisms compete for votes in simulation. Later iterations obtain monotonically increasing votes against earlier versions, with convergence after iteration 7. The initial iteration (random mechanism) is denoted with $i$.}
    \label{fig:improvements}
\end{figure*}

Related to the choice of training updates within an iteration's \texttt{OPTIMIZE} step, our method requires determining how many {\it iterations}, i.e. repetitions of our \texttt{ACQUIRE}, \texttt{MODEL} and \texttt{OPTIMIZE} pipeline, we should complete. Our proposed approach is to construct a meta-game: a two-player normal form game with payoff matrix of size $s \times s$ and entries $i,j$ corresponding to the proportion of votes collected by mechanisms playing with parameters $\theta_m^i$ and $\theta_m^j$ over $100$ repetitions of the curried game $\mathcal{G}^{\bar{\theta}_p^{s}}$ constructed using the latest participants' models (see Fig.~\ref{fig:improvements}). Once the actions corresponding to later checkpoints no longer constitute a dominant strategy in the meta-game, or when their advantage becomes negligible, we conclude that HCMD-{\it zero} has converged, since it no longer produces meaningful improvements.

\begin{figure*}[h]
    \centering
    \includegraphics[width=\linewidth]{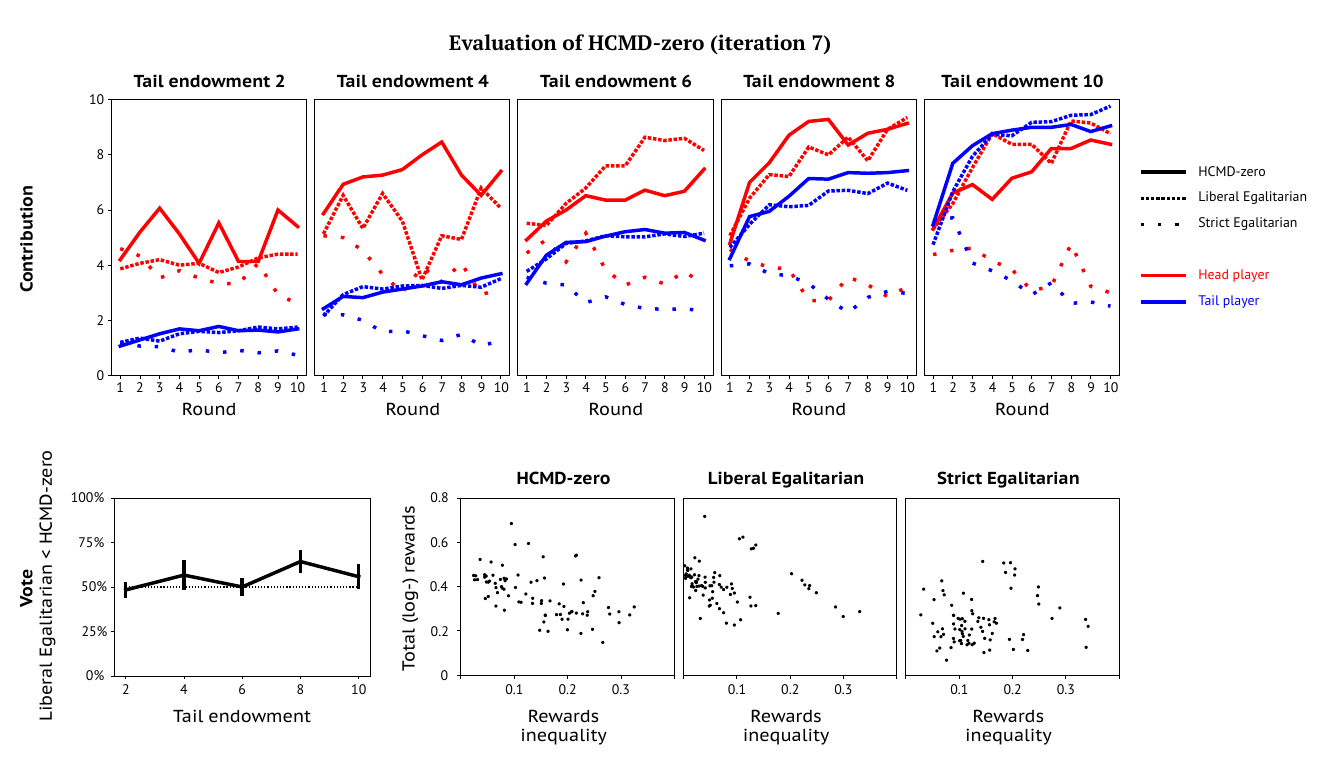}
    \caption{\small Evaluation of HCMD-{\it zero} after convergence against Liberal Egalitarian. Top panel: contribution of head and tail participants across rounds, as a function of endowment, for each mechanism. Bottom left panel: votes in favour of HCMD-{\it zero} against Liberal Egalitarian, as a function of tail endowment. This was the metric the mechanism aimed to maximize. Bottom-right panel: Scatter plot of total reward (sum of log-rewards) against reward inequality (Gini coefficient), for each mechanism. Each dot corresponds to one group, aggregated across all endowments.}
    \label{fig:evaluation}
\end{figure*}

\section{Results}
\begin{wrapfigure}{R}{6cm}
    \centering
    \includegraphics[width=7cm]{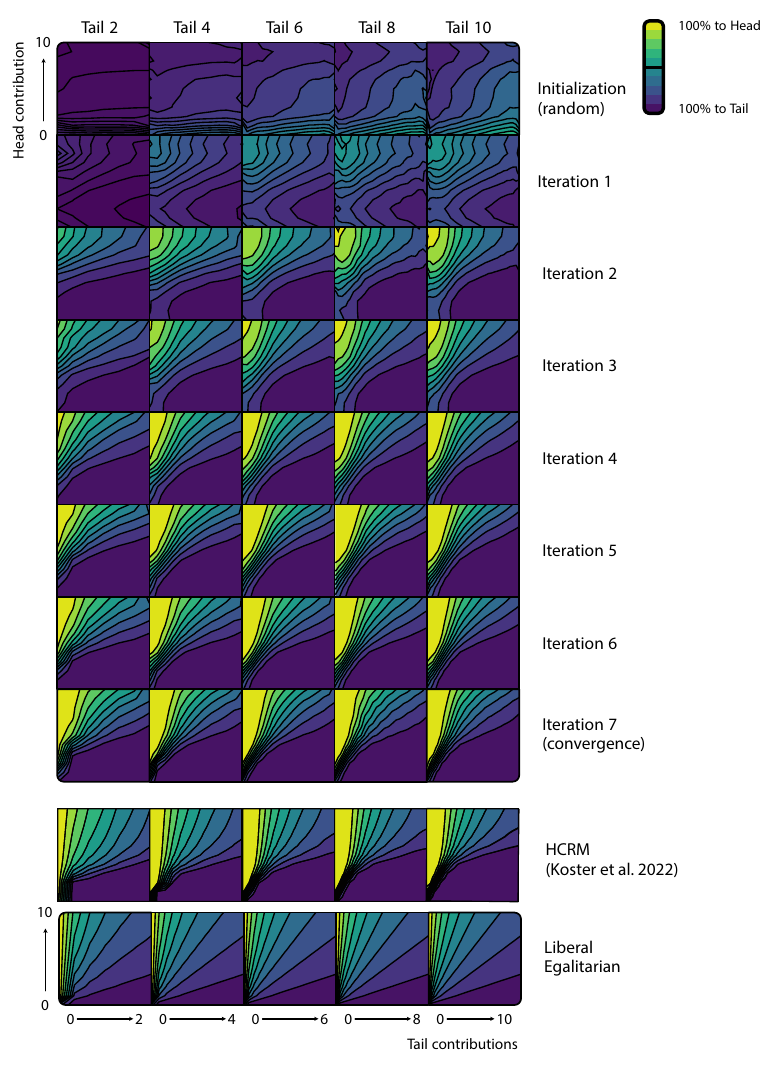}
    \caption{\small Illustration of the learnt mechanism policy across training iterations. Each heat-map illustrates a mechanism (rows) under a given tail endowment (columns). The rows correspond to mechanism $\theta_m^{0}$ followed by iterations $\theta_m^s$. The policy becomes more similar the one reported in~\cite{koster2022democratic}. The last row corresponds to the competing mechanism used in evaluation. Heat-maps illustrate the fraction of payout distributed to the head (yellow) or tail participants (blue) as a function of the contributions provided by the head (y-axis, bottom to top) and tail participants (x-axis, left to right).}
    \label{fig:beachplots}
\end{wrapfigure} 

In this section we show the results of applying our method in the Public Investment Game. We first show the performance against baselines after 7 iterations of training with HCMD-{\it zero}. Then, we explore in more detail the learning dynamics of the model of human participants, as well as the convergence of the mechanism. Finally, we provide an analysis of the mechanism's behavior throughout training.

\subsection{Performance of HCMD-{\it zero} at evaluation}
In order to validate our approach, we trained a mechanism in the Public Investment Game (PIG). Similar to ~\cite{koster2022democratic}, we divided participants into one ``Head'' participant that always received an endowment $e_{head,t} = 10$ coins and three ``tail'' participants that received a ``tail'' endowment $e_{tail,t} \in \{2, 4, 6, 8, 10\}$ coins. The tail endowment was consistent within a group, across tail payers and for all Stages if the game (i.e. for all the mechanisms they interacted with). We evaluated the mechanism obtained after 7 iterations $\theta_m^{7}$ by collecting new data specifically for this purpose. Humans interacted with the trained mechanism and with a baseline alternative in two subsequent games in counterbalanced order (Stages 1 and 2). Our choice of baseline was the Liberal Egalitarian mechanism, a redistribution scheme that disburses the public fund according to the proportion of endowment contributed by each participant, as baseline. Koster et al. \cite{koster2022democratic} show that Liberal Egalitarian is a strong baseline that is preferred by humans over the Strict Egalitarian, which divides the fund in equal parts.

Results are shown in Fig.~\ref{fig:evaluation}. HCMD-{\it zero} was voted more often than Liberal Egalitarian, achieving an average of 54.3\% of the votes ($p < 0.06$ with a non-parametric analysis that corrects for in-group correlations). More specifically, HCMD-{\it zero} achieved at least half of the votes against Liberal Egalitarian (see bottom-left panel), whilst matching the contributions from participants (top panel). At the group level, HCMD-{\it zero} matched the performance of Liberal Egalitarian in trading off the productivity of the group (incentivizing the head participant to contribute more) and the inequality of the group (redistributing to the tail participants; bottom-right panel).

\subsection{Participant models display behavior shifts}
We turned to look at the predictive power of the participant model $\bar{\theta}_p^s$ across iterations. Our iterative method addresses the fact that human contribution and voting behavior depends on the mechanism. Fig.~\ref{fig:improvements} (left two panels) shows that this effect is observed in practice. We construct a contribution and vote cross-validation matrix by reporting in entry $i,j$ the cross-entropy loss achieved by each model $\bar{\theta}_p^i$ (rows) on each data-set $D_j$ (columns); recall that model $\bar{\theta}_p^i$ is trained using data-sets $D_1,\ldots,D_i$ (matrix entries are normalized per-column by the corresponding diagonal entry). The figure clearly shows that the predictive performance of each model degrades progressively for each subsequent data-set indicating that participants contributing and voting behavior has changed. For example, the normalized contribution cross-entropy for data-set $D_4$ was 23.3\% for Participant model $f_p(\bar{\theta}_p^2)$ and 13.8\% for Participant model $f_p(\bar{\theta}_p^3)$ indicating a larger distributional shift between $D_2$ and $D_4$ than between $D_2$ and $D_3$ (note that lower is better here, as we are comparing losses).

\subsection{Mechanism improvement and convergence}
On every iteration $s$ we constructed a meta-game as described in the methods above, where each row and column corresponds to the mechanisms $\theta_m^0, \ldots, \theta_m^s$ and each cell corresponds to the number of votes obtained in simulation with the participant model $\bar{\theta}_p^s$. This can be found for iteration 7 on the right panel in Fig.~\ref{fig:improvements}. With HCMD-{\it zero}, the optimization of the mechanism showed consistent improvements on every iteration, with diminishing returns until convergence on iteration 7.
This method achieves a policy similar to \cite{koster2022democratic} (see Fig.~\ref{fig:beachplots}), while using only about a quarter of the data (1242 vs 4809 individual games of 10 rounds, but note that the latter contains some games where participants dropped out and their responses were replaced with random actions).

\subsection{Analysis of mechanism behavior}
We analyze the learnt mechanism policy across iterations in Fig.~\ref{fig:beachplots}. For each tail endowment (columns) and across iterations (rows), we illustrate the mechanism's policy on a grid containing the contributions of the head (y-axis) and tail participants (x-axis, averaged across the 3 tail participants). Then, for each possible contribution pair, we computed the average redistribution weight across episodes and players. These are plotted with yellow favouring redistribution to the head participant (high endowment) and blue to the tail participants (low endowment). For example, Liberal Egalitarian (see most bottom row) displays straight lines fanning out from left to right and bottom to top (e.g. the yellow region indicates where the head participant gives the most and in turn receives the most back). 
Over iterations, our mechanism learns a policy similar to the mechanism reported in \cite{koster2022democratic}, despite being exposed to less data and not having access to strong baselines during training. For example for the tail 8 condition, the redistribution to head for not contributing when tail players only contributed 2 coins was about 17\% at iteration 4, whereas at iteration 7 it was 6.1\% indicating the mechanism punished the head player more harshly for contributing a low fraction of its endowment in the later iteration. Like Liberal Egalitarian, these two trained mechanisms set clear incentives to contribute, but also redistributes the wealth created to the tail participants. However, they expand the areas in which low contributions are under-rewarded, creating a ReLU-like function. This policy is interpretable as a threshold below which any contributions are considered defection and get no returns from the pool.

\section{Discussion}

We have introduced HCMD-{\it zero}, a general purpose method to construct mechanisms that are shaped and eventually preferred by the population of human participants that interact with. HCMD-\textit{zero} requires no baseline or alternative mechanisms during training, and no knowledge of the environment dynamics. It also makes no assumptions about what a desirable policy for the task is from the view of the human participants. Our method uses participant modeling and self-play to minimize the amount of data that is required to train a mechanism, and it iteratively addresses the challenges posed by behavior shifts, where the participants behavior changes in response to updates in the mechanism policy.
Our results show that HCMD-{\it zero} produces a competent mechanism player in the Public Investment Game, a game that exposes multiple social conflicts and allows many possible solutions. Our detailed analysis shows that our mechanism policy is consistently improved across iterations, and provides an interpretation of its final policy.

Minimizing assumptions and prior knowledge also exposes important limitations. Letting a policy be shaped by the population it interacts with is only as good as the democratic process allows (e.g. dictatorship of the majority or strategic voting).
Another limitation is that mechanisms would ideally be interpretable to the population whose interactions it stand to mediate. This is a challenge for complex neural networks, and an analysis of the policy like we provide here would not have been possible if the network had been furnished with memory.
Perhaps most importantly, our method in the form presented here, is only viable in scenarios where it is safe to explore and gather data under unfinished policies. If applied to real world problems, it may be crucial to restrict the policy space a priori to avoid gathering data under a generally harmful policy (e.g. with a warm start on simulated data).

Artificial learning agents are becoming a centerpiece of our institutions, and as such methods to ensure that mechanisms are aligned to the values of their constituents are being heavily investigated. The ideas and results presented here indicate that integrating the most basic democratic principle of one person one vote, with modern machine learning techniques can be viable and fruitful path forward.

\bibliographystyle{unsrt}
\bibliography{main}
\end{document}